\documentclass[a4paper]{article}

\usepackage{cmap}
\usepackage[utf8]{inputenc}
\usepackage[T1]{fontenc}
\usepackage[english]{babel}

\usepackage{amsmath,mathtools,amssymb,amsfonts,bm,bbm,amsthm}
\allowdisplaybreaks
\theoremstyle{plain}
\newtheorem{theorem}{Theorem}
\newtheorem{lemma}{Lemma}
\newtheorem{proposition}{Proposition}

\theoremstyle{definition}
\newtheorem{definition}{Definition}
\newtheorem{assumption}{Assumption}

\newtheorem{remark}{Remark}

\usepackage[bottom=3cm, margin=1in]{geometry}
\usepackage{times}
\usepackage{microtype}

\usepackage[parfill]{parskip}

\usepackage{titling}
\usepackage[]{titlesec}
\usepackage{needspace}

\usepackage{float}
\usepackage[font={small}, hang]{caption}

\usepackage{tikz}
\usetikzlibrary{graphs}
\usetikzlibrary{shapes.symbols, decorations.pathmorphing, fadings, shapes}
\usetikzlibrary{arrows}
\tikzset{node distance=2cm,, >=stealth', every picture/.style={line width=0.6pt}}

\usepackage{graphicx} 
\usepackage{xcolor}

\usepackage[style=apa, natbib]{biblatex}
\usepackage[autostyle]{csquotes}
\usepackage[hidelinks, pdfpagelabels]{hyperref}
\hypersetup{
	pdfborder={0 0 0.5},
	linkcolor={blue!50!black},
	citecolor={blue!50!black},
	citebordercolor={1 0 0},
	urlcolor={blue!50!black},
	colorlinks=true
}
\usepackage[all]{hypcap}
\usepackage{cleveref}
\Crefname{assumption}{Assumption}{Assumptions}

\renewcommand*\d{\mathop{}\!\mathrm{d}}
\newcommand{\ind}{\perp \mkern-10mu \perp}
\renewcommand{\bar}[1]{\mkern 1.5mu\overline{\mkern-1.5mu#1\mkern-0.5mu}\mkern 1.5mu} 
\newcommand{\e}{\mathrm{e}}
\usepackage{prodint}
\usepackage{nicefrac}  

\usepackage{authblk} 
\title{On the limitations of causal inference with current-treatment Cox models}
\date{}
\author[1]{Mark B. Knudsen}
\author[1]{Erin E. Gabriel}
\author[1]{Torben Martinussen}
\author[1]{Helene C. W. Rytgaard}
\author[2]{Arvid Sj{\"o}lander}
\affil[1]{Section of Biostatistics, Department of Public Health, University of Copenhagen, Copenhagen, Denmark}
\affil[2]{Department of Medical Epidemiology and Biostatistics, Karolinska Institute, Stockholm, Sweden}

\begin{document}

\maketitle

\begin{abstract}
	Cox models with time-varying treatments often include only the current treatment level. Translating such a model into causally meaningful intervention-specific survival probabilities relies on the Markov property: that the hazard is independent of treatment history conditional on current treatment. For the Markov property to not be population-specific, it needs to hold also conditional on any unmeasured prognostic heterogeneity (frailty). Using a discrete-time argument, earlier work concluded that the Markov property can hold both conditionally and marginally only in the absence of a treatment effect or an effect of the unmeasured heterogeneity. We broaden this argument by developing a continuous-time framework that encompasses both proposed extensions of the Kaplan-Meier curve and current-treatment marginal structural Cox models, and sharpen it by characterizing precisely the conditions under which the Markov property can hold both conditionally and marginally. Specifically, we show that it requires the absence of treatment-frailty interaction on the additive hazard scale. As frailty is inherently unmeasured, such a no-interaction assumption cannot be verified. Consequently, causal interpretation of a current-treatment marginal structural Cox model rests on a strong and unverifiable structural assumption. Illustrating this point, we construct a setting in which a Cox model is correctly specified conditionally on current treatment only, although the marginal Markov property fails. Transforming the fitted model to the survival scale does not recover the true intervention-specific survival probabilities. Thus, moving from the hazard scale to the survival scale is not in itself sufficient to obtain a causal interpretation in the time-varying treatment setting.
\end{abstract}


\section{Introduction}

Time-to-event methods are commonplace in medical research, and the Cox model \citep{coxRegressionModelsLifeTables1972} is by far the most commonly used regression model among them. The Cox model parametrizes the hazard function, which is the instantaneous rate of events occurring among subjects who have not yet experienced the event. The hazard ratio (HR) comparing treated and untreated is often the primary effect measure reported in studies with a time-to-event outcome. In recent years, however, warnings have been issued about the subtle causal interpretation of the HR, as it mixes a (potentially) real treatment effect with an unavoidable selection effect if the treatment actually has a causal effect \citep{hernanHazardsHazardRatios2010, aalenDoesCoxAnalysis2015, martinussenSubtletiesInterpretationHazard2020}. Due to this, \citet{hernanHazardsHazardRatios2010} recommends supplementing HRs with estimated survival curves for easier causal interpretation. Many recent contributions to the causal time-to-event literature likewise focus on non-hazard estimands such as the probability of survival/death or the restricted mean survival time. In the Cox model with baseline binary treatment, the HR parameter equals the ratio of the log-survival probabilities for the treated and untreated, yielding a probability-based causal interpretation of the HR parameter in the absence of unmeasured confounding \citep{martinussenSubtletiesInterpretationHazard2020}. This interpretation is, however, less straightforward.

When treatment is time-varying, Cox models are commonly specified using only the current treatment level as a time-dependent covariate. In order to be a model for the hazard of a potential outcome under a hypothetical intervention on the treatment process, the Markov property needs to be fulfilled: conditional on current treatment, the hazard should be independent of past treatment. The scientific justification for such a marginal Markov property is less straightforward. A mechanistic explanation would naturally be formulated at the individual level, conditional on prognostic characteristics, including unmeasured frailty that affects the outcome but is independent of treatment. For this explanation to support a marginal Markov property, the conditional Markov property must be preserved after marginalizing over this heterogeneity. If conditional Markov fails, a marginal Markov property can only arise through cancellation after averaging over frailty. Its validity then depends on the distribution of frailty in the study population and cannot generally be transported to another population. The difficulty of reconciling the conditional and marginal Markov properties was highlighted by \citet{sjolanderCautionaryNoteExtended2020} who, in critically examining the causal interpretation of the extended Kaplan-Meier curve of \citet{snapinnIllustratingImpactTimeVarying2005, simonNonparametricGraphicalRepresentation1984}, used a discrete-time approach to show that the conditional and marginal Markov property can hold simultaneously only in the absence of an effect of either the treatment or frailty. The former leaves no treatment effect to interpret, whereas the latter rules out any unmeasured prognostic heterogeneity and is therefore implausible in realistic applications.

We revisit this problem in continuous time and broaden \citeauthor{sjolanderCautionaryNoteExtended2020}'s argument by developing a general framework that encompasses proposed Kaplan–Meier extensions, current-treatment Cox models, marginal structural Cox models (Cox MSMs), and multi-state models with an absorbing state. We sharpen the critique by characterizing precisely the conditions under which the Markov property can hold both conditionally and marginally in the continuous-time setting. Specifically, we show that it requires the absence of treatment-frailty interaction on the hazard scale: the frailty-conditional hazard must be additively separable into a frailty component and a treatment component. Frailty may be independent of the treatment process and therefore need not constitute unmeasured confounding. Nevertheless, because it is unmeasured, the required no-interaction condition cannot be verified from the observed data. The result has direct implications for Cox MSMs that include only current treatment \citep{hernanMarginalStructuralModels2000, hernanMarginalStructuralModels2001}, a specification commonly used in applied research \citep{chirgwinTreatmentAdherenceIts2016, choiMethotrexateMortalityPatients2002, liAssociationReligiousService2016, nahonIncidenceHepatocellularCarcinoma2018, tiihonen11yearFollowupMortality2009}. As in the baseline treatment setting, the resulting parameter has no simple causal interpretation as a hazard ratio. Translating the fitted model into intervention-specific survival probabilities does not solve the problem on its own, as it requires the hazard to be correctly specified and therefore relies, under a current-treatment specification, on the Markov property and the associated unverifiable no-interaction condition. The problem therefore concerns the structural assumptions required for causal interpretation, rather than merely the choice of effect scale.

The idea that unmeasured frailty can induce dependence on the history of a process, and thus a violation of the Markov property, has been explored previously by \citet{putterFrailtiesMultistateModels2015} in the context of multi-state models. Some works suggest including a shared multiplicative frailty term in all transition hazards of an illness-death model to account for residual correlation between transition times \citep{haneuseSemiCompetingRisksData2016}. However, these works focus on the feasibility and usefulness of frailty models in explaining how a hazard depends on the history. In this work, we view the presence of frailty as a more fundamental issue for the validity of the Markov property.

If the Markov property is not satisfied, the current-treatment Cox model may still be correctly specified as a model for the rate, defined as the instantaneous risk of an event conditional on survival and current treatment only. However, this rate has no clear causal interpretation, except if the Markov property holds, in which case it equals the hazard conditional on the entire treatment history. We show that the extended Kaplan–Meier curve estimates the exponential of minus the cumulative rate, thereby confirming that a causal interpretation relies on the Markov property being true. We further construct an example in which the current-treatment Cox model is correctly specified for the rate, although the Markov property fails. Consequently, the survival probabilities derived from the fitted model, targeting the same quantity as the extended Kaplan-Meier curve, differ from the true intervention-specific survival probabilities. 

The remainder of the paper is organized as follows. In \Cref{sec:setup} we introduce the hazard- and rate functions, define the Markov property, and explain its connection to Cox MSMs. In \Cref{sec:markov_problem}, we characterize the conditions under which the Markov property can hold both marginally and conditional on frailty, and establish the requirement of no treatment–frailty interaction on the hazard scale. In \Cref{sec:causal_rate} we show that the extended Kaplan-Meier curve estimates the exponential of minus the cumulative rate. In \Cref{sec:example}, we provide an example of a proportional rates model in which the Markov property does not hold and show that using a Cox MSM in this setting yields incorrect estimates of the intervention-specific survival probabilities. Finally, in \Cref{sec:discussion}, we discuss the broader implications of our results. Technical derivations are deferred to the Appendix.

\section{Notation and preliminaries}
\label{sec:setup}

This section collects notation and key preliminaries. While the primary focus of this paper is a time-varying treatment setting with a time-to-event outcome, we first consider the simpler setting with a baseline treatment to provide context. Here we review the arguments that justify a certain causal interpretation of the hazard ratio under standard identifiability assumptions. We then move on to the time-varying treatment setting, where we carefully distinguish between hazards and rates and highlight the role of the Markov assumption. Note that we focus throughout on data settings without covariates (baseline or time-varying), and that we generally assume no confounding. We emphasize that introducing confounding would add further complications, and that the issues we point out persist even without the presence of confounding.

\subsection{The Cox model for baseline treatment}
\label{subsec:simpel_setup}

Consider first the setting with a baseline treatment $A$, where interest is in assessing how this affects a time-to-event outcome $T$. The \emph{hazard function}
\begin{equation*}
	\lambda(t \mid a) = \lim\limits_{\Delta t \to 0} \frac{1}{\Delta t} P(T \in [t, t + \Delta t) \mid T \geq t, A = a)
\end{equation*}
uniquely determines the conditional distribution of $T$ given $A$ and is the target of most time-to-event regression models. The most commonly used regression model is Cox' proportional hazards model \citep{coxRegressionModelsLifeTables1972}
\begin{equation}
\label{eq:Cox_hazard_baseline}
    \lambda(t \mid a) = \lambda_0(t) \e^{\beta a},
\end{equation}
where $\lambda_0(t)$ is some unspecified baseline hazard function and $\beta \in \mathbb{R}$ a regression parameter. The ratio between the hazards for two different values of $A$ is assumed constant over time, and when $A$ is binary the HR comparing the two groups is $\e^{\beta}$.

The ambition of most studies is to learn about the causal effect of a treatment or exposure on an outcome. For a binary baseline treatment, a causal effect is a contrast between the distributions of $T^0$ and $T^1$, where $T^a$ is the potential outcome under treatment $a$. The Cox model
\begin{equation}
\label{eq:cox_baseline_causal}
    \lambda^{a}(t) = \lim\limits_{\Delta t \to 0} \frac{1}{\Delta t} P(T^a \in [t, t + \Delta t) \mid T^a \geq t) = \lambda_0(t) \e^{\beta a}
\end{equation}
for the hazard $\lambda^{a}(t)$ of $T^{a}$ is known as a marginal structural Cox model (Cox MSM) \citep{robinsMarginalStructuralModels2000a, hernanMarginalStructuralModels2000, hernanMarginalStructuralModels2001}. Even though the HR, $\e^{\beta}$, relates to the potential outcomes, its causal interpretation has been brought into question. We return to this point below. We assume the standard identification conditions:
\begin{enumerate}
    \item Consistency: the observed outcome equals the potential outcome corresponding to the received treatment, $T = T^A$, and 
    \item Exchangeability: the potential outcomes are independent of assigned treatment, $T^a \ind A$ for $a \in \{0, 1\}$, as would be the case in a randomized study, and
    \item Positivity: $P(A = a) > 0$ for $a \in \{0, 1\}$.
\end{enumerate}
Under these conditions, the potential outcome hazard $\lambda^a(t)$ equals the observed $\lambda(t \mid a)$ and the HR can be estimated by fitting a Cox model to observed data with treatment as a covariate. This is standard practice in clinical trials with time-to-event endpoints.

However, as highlighted by \citet{hernanHazardsHazardRatios2010}, hazard contrasts like the HR are difficult to interpret causally, even when treatment $A$ is randomized at baseline. In the presence of unmeasured frailty (a cause of the outcome, independent of treatment), a non-zero effect of treatment will induce a selection effect over time: if the treatment is protective, more frail subjects will survive to time $t$ with treatment than without treatment. The populations $(T^1 \geq t)$ and $(T^0 \geq t)$ that are being conditioned on in the potential outcome hazards \eqref{eq:cox_baseline_causal} are therefore not directly comparable for $t > 0$. Even if the effect of treatment is really constant over time for subjects of equal frailty, the increasing frailty imbalance would cause the observed HR to be time-varying, potentially decaying to 1. Such a time-varying HR has often been interpreted as an attenuation of the treatment effect, but it could also be caused purely by selection effects. Causal contrasts based on marginal quantities, such as survival probabilities or restricted mean survival times, avoid the issues mentioned above as they do not condition on survival past time 0. Particularly, \citet{martinussenSubtletiesInterpretationHazard2020} point out that the Cox model \eqref{eq:cox_baseline_causal} for baseline treatment is equivalent to assuming that
\begin{equation*}
\label{eq:log_surv_ratio_baseline}
    \frac{\log(P(T^1 > t))}{\log(P(T^0 > t))} = \e^{\beta}
\end{equation*}
for all $t > 0$, such that the HR parameter can instead be interpreted as the ratio of the log-survival functions of the potential outcomes. This yields a probability-based interpretation of the regression parameter which is free from the issues described above.

\subsection{The Cox model with time-varying treatment}
\label{subsec:time-varying}

For the remainder of the paper we consider the setting where the treatment/exposure can vary over time, represented by a process $A(t)$. Examples of this includes drug usage, a surgery taking place post-baseline, or air pollution levels. This introduces some ambiguity in how the hazard function should be defined, as it is possible to condition on different parts of the history of $A(t)$. Hazard models are most often introduced as \emph{intensity models}, where the hazard is defined conditional on the entire treatment history $\bar{A}(t) = \{A(s) \mid s < t\}$. We make the dependence on $\bar{A}(t)$ explicit in the notation and define the \emph{hazard} as
\begin{equation}
	\label{eq:fully_conditional_hazard}
	\lambda(t \mid \bar{a}(t)) = \lim\limits_{\Delta t \to 0} \frac{1}{\Delta t} P(T \in [t, t + \Delta t) \mid T \geq t, \bar{A}(t) = \bar{a}(t)).
\end{equation}
The dependence on the history $\bar{a}(t)$ can be complex, making modeling of the hazard a difficult task without further assumptions. The \emph{Markov property} is exactly such an assumption, stating that the hazard is only influenced by current treatment. More precisely, the Markov property asserts that $\lambda(t \mid \bar{a}(t)) = r(t \mid a(t-))$, where the latter quantity is defined as
\begin{equation}
	\label{eq:partly_conditional_hazard}
	r(t \mid a) = \lim\limits_{\Delta t \to 0} \frac{1}{\Delta t} P(T \in [t, t + \Delta t) \mid T \geq t, A(t-) = a),
\end{equation}
where $A(t-)$ denotes the left-hand limit of $A(t)$. The only difference between $\lambda(t \mid \bar{a}(t))$ and $r(t \mid a)$ is the amount of treatment history they condition on. Thus, $r(t \mid a)$ can also be thought of as a type of hazard, but given different information. A similar quantity has been studied in the context of recurrent events and longitudinal data, where it has been referred to as the \emph{rate function} \citep{linSemiparametricRegressionMean2000} or the \emph{partly conditional rate} \citep{pepeModelingPartlyConditional1997}. \citet[section 6.5]{kalbfleischStatisticalAnalysisFailure2002} describe this quantity in a survival time context. We will refer to $r(t \mid a)$ as the rate function, hence making a distinction between the hazard and the rate. The rate can be obtained from the hazard via an iterated expectation
\begin{equation}
\label{eq:rate_from_hazard}
    r(t \mid a) = E(\lambda(t \mid \bar{A}(t)) \mid T \geq t, A(t-) = a),
\end{equation}
such that the rate is defined even in non-Markov settings, in which case it does not equal the full hazard function. Note that if treatment is fixed at baseline, then there is no difference between the hazard and the rate. When defining the rate, one could have chosen to condition on other summaries of $\bar{A}(t)$ than just the current value $A(t-)$, yielding a different rate function for every possible summary, each one corresponding to a type of Markov property if assumed to be equal to the hazard. In that sense, the Markov property defined above is the most extreme out of many possible assumptions about independence of past treatment. 

The Cox model with a time-varying treatment is typically obtained from \eqref{eq:Cox_hazard_baseline} by substituting $a(t-)$ in place of $a$. The question is whether this is supposed to be a model for the full hazard \eqref{eq:fully_conditional_hazard} or the rate \eqref{eq:partly_conditional_hazard}. Sources are not always explicit about this, but mathematically precise treatments (such as \citet{andersenStatisticalModelsBased1993, andersenCoxsRegressionModel1982}) typically model the hazard
\begin{equation}
	\label{eq:cox_hazard}
	\lambda(t \mid \bar{a}(t)) = \lambda_0(t) \e^{\beta a(t-)},
\end{equation}
thereby assuming the Markov property. This is because the full hazard is closely connected to the intensity process in martingale theory, which is used to establish asymptotic properties of estimators. It is also possible to assume a Cox structure for the rate function, such that
\begin{equation}
	\label{eq:cox_hazard_partly}
	r(t \mid a(t-)) = r_0(t)  \e^{\beta a(t-)},
\end{equation}
where $r_0(t)$ is some unspecified baseline rate. The solution $\hat{\beta}$ to Cox's score equation \citep{coxRegressionModelsLifeTables1972} is a consistent estimator of $\beta$ in either model (when correctly specified). As a side remark, we note that the usual martingale-based variance estimator, used in the hazard model \eqref{eq:cox_hazard}, is not correct in the rate model. Instead a robust sandwich estimator must be used \citep{linSemiparametricRegressionMean2000, kalbfleischStatisticalAnalysisFailure2002}. Since the rate does not condition on the past history, the Cox rate model \eqref{eq:cox_hazard_partly} does not assume the Markov property.

\subsection{Marginal structural Cox models}
\label{sec:causal_cox}

For a time-varying treatment, each possible trajectory $\bar{a}$ has an associated potential outcome $T^{\bar{a}}$ that would have been observed, had treatment been forced to follow $\bar{a}$. We are particularly interested in the cases of ``never-treat'' and ``always-treat'', which we denote as $\bar{0}$ and $\bar{1}$, respectively. For the hazard $\lambda^{\bar{a}}(t)$ of the potential outcome $T^{\bar{a}}$ it is common to assume a Cox MSM including only the treatment $a(t-)$ currently prescribed by the strategy $\bar{a}$ \citep{hernanMarginalStructuralModels2000, hernanMarginalStructuralModels2001}:
\begin{equation}
\label{eq:cox_timevar_causal}
    \lambda^{\bar{a}}(t) = \lambda_0(t) \e^{\beta a(t-)}.
\end{equation}
Time-varying treatments are most commonly seen in observational studies, where it is necessary to control for time-varying confounders. In the MSM framework, this is accomplished by fitting a Cox model with a time-varying treatment to observed data weighted by time-varying inverse probability of treatment weights (IPTW) and assuming sequential exchangeability conditional on the confounders \citep{hernanMarginalStructuralModels2000, hernanMarginalStructuralModels2001}. In order to focus our discussion we will assume that there are no confounders, in which case $\lambda^{\bar{a}}(t) = \lambda(t \mid \bar{a}(t))$. Then the parameter $\e^\beta$ in the MSM \eqref{eq:cox_timevar_causal} can be estimated by fitting a Cox model \eqref{eq:cox_hazard} to observed data. 

When viewed as a hazard ratio, the parameter $\e^\beta$ suffers from the same interpretational difficulties as in the baseline treatment setting. Similarly to before, it can be expressed as the ratio between the log-survival functions of the potential outcomes under the always-treat and never-treat strategies
\begin{equation}
\label{eq:log_surv_ratio_timevar_markov}
    \frac{\log(P(T^{\bar{1}} > t))}{\log(P(T^{\bar{0}} > t))} = \e^{\beta},
\end{equation}
resulting in a more direct causal interpretation. This representation relies on \eqref{eq:cox_timevar_causal} being true, and so depends crucially on the Markov property. 

\section{Conditional and marginal Markov properties}
\label{sec:markov_problem}

In a discrete-time setting, \citet{sjolanderCautionaryNoteExtended2020} argued that in the presence of a \emph{frailty variable} $Z$, i.e. an unmeasured cause of the outcome that is independent of the exposure, the Markov property essentially cannot hold both conditionally on $Z$ and marginally (not conditional on $Z$) at the same time, and therefore the marginal version is implausible. We here give corresponding results for the continuous-time setting, where the picture is more nuanced. Indeed, we show that the Markov property can hold both conditionally and marginally, but only in the absence of effect modification by $Z$ on the hazard scale.

First, it is worth considering why we should be interested in whether the Markov property holds conditional on some unmeasured frailty variable. Indeed, a Cox MSM, or any other method relying on the Markov property, only needs it to hold for the observed data. There are multiple reasons for this. Firstly, researchers applying Markov methods rarely attempt to justify the assumption using scientific background knowledge. However, in a setting where the analyst really does believe in the Markov property, one would expect there to be some etiological reasoning behind this, and any such explanation will necessarily be mechanistic and therefore conditional in nature. Consider an example where the exposure is some drug which patients consume daily while under treatment. There might be pharmacodynamic reasons to believe that the drug only has an effect via the current concentration in the bloodstream, tempting one to conclude that a current-treatment model is warranted. However, such a mechanistic explanation applies at the biological level, i.e. conditional on all biological factors $Z$ affecting the outcome, and therefore supports a conditional rather than a marginal Markov property. Secondly, as we show below, if the conditional Markov property is not satisfied, then the marginal Markov property can hold only in a single population, as determined by the frailty distribution. Thus, in order for the Markov property to be a substantive assumption about the world rather than a population-specific attribute, the conditional version must hold. For these reasons, it is of interest to consider the conditions under which a conditional Markov property implies a marginal one, as these would have to be satisfied in order for the analyst to have substantive justification for their use of a current-treatment model.

To explain why one should generally not expect the Markov property to hold both conditionally and marginally at the same time, we first briefly review the discrete-time argument \citep{sjolanderCautionaryNoteExtended2020}. Suppose that exposure- and event status is only updated every $\delta > 0$ time units, and denote these values by $A_t$ and $N_t$. We assume that $A_t$ can affect $N_t$ but not vice-versa. Consider now a time $t$ and the following one $t + \delta$. The variables $Z$, $A_t$, $N_t$, $A_{t + \delta}$, $N_{t + \delta}$ can be described by the causal directed acyclic graph (DAG) on \Cref{figure:sjolander_problem}. We refer to e.g. \citet[chapter 6]{hernanCausalInferenceWhat2020} for details on causal DAGs. The omission of the arrow $A_t \to N_{t + \delta}$ is due to the conditional Markov property which states that $N_{t + \delta} \ind A_t \mid N_t = 0, A_{t + \delta}, Z$, and the lack of direct arrows from $Z$ to $A_t$ and $A_{t + \delta}$ is the assumption of independence between frailty and exposure. Now, conditioning on survival up to time $t$ opens the path $A_t \rightarrow N_t \leftarrow Z \rightarrow N_{t + \delta}$ by conditioning on the collider $N_t$, and we conclude that $N_{t + \delta} \not\ind A_t \mid N_t = 0, A_{t + \delta}$. Note that the last step requires assuming that the distribution is \emph{faithful} to the DAG \citep{zhangThreeFacesFaithfulness2016}, which is natural in most settings. Thus, the Markov property cannot hold marginally without a violation of faithfulness, unless the exposure or the frailty has no effect.

\begin{figure}
	\centering
	\begin{tikzpicture}
        \tikzset{node distance=1.5cm}
        \node (Z1) {$A_t$};
        \node (D1)[shape=rectangle, draw] [below right of=Z1] {$N_t$};
        \node (Z2) [above right of=D1] {$A_{t + \delta}$};
        \node (D2) [below right of=Z2]{$N_{t + \delta}$};
        \node (eta) [below left of=D1] {$Z$};
        
        \graph{
            (Z1) -> (Z2),
            (Z1) ->[dashed] (D1),
            (Z2) -> (D2),
            (D1) -> {(Z2), (D2)},
            (eta) ->[dashed] {(D1), (D2)}
        };
    \end{tikzpicture}
	\caption{A directed acyclic graph where the exposure $A_t$ has no direct effect on the outcome $N_{t + \delta}$, such that the Markov property is satisfied conditional on $Z$. Dashed arrows indicate a path between $A_t$ and $N_{t + \delta}$ which is opened up by conditioning on $N_t = 0$, implying that the Markov property is not satisfied marginally (not conditional on $Z$).}
	\label{figure:sjolander_problem}
\end{figure}

Ultimately, the violation of the marginal Markov property in discrete time happens because the survivor frailty distribution $Z \mid N_t = 0, A_t, A_{t + \delta}$ depends on $A_t$. Interestingly, this dependence need not arise in continuous time. Here, marginalizing over $Z$ preserves the Markov property if there is no treatment-frailty interaction on the hazard scale. To see this, consider a setting with baseline exposure $A$ independent of $Z$. Here it is known that if the hazard $\lambda(t \mid a, z)$ of $T$ given $A = a$ and $Z = z$ has an additive structure $\lambda(t \mid a, z) = h_A(t, a) + h_Z(t, z)$ then $Z \ind A \mid T > t$ for all $t > 0$ \citep{aalenDoesCoxAnalysis2015}, and so the survivor frailty distribution does not depend on the exposure at all. An analogous result can be shown for a time-varying exposure $A(t)$ which is locally independent of $Z$ (see \Cref{sec:local_independence} for a definition). We show in \Cref{result:Z_distribution_survivor} in \Cref{sec:frailty_survivor} that the distribution of $Z \mid T \geq t, \bar{A}(t) = \bar{a}(t)$ has density
\begin{equation*}
    p_t(z \mid \bar{a}(t)) = \frac{\exp\left( - \Lambda(t \mid \bar{a}(t), z) \right) p(z)}{\int \exp\left( - \Lambda(t \mid \bar{a}(t), z) \right) p(z) \d z},
\end{equation*}
where $p(z)$ denotes the density of $Z$. If the hazard function $\lambda(t \mid \bar{a}(t), z)$ of $T$ given $\bar{A}(t) = \bar{a}(t)$ and $Z = z$ has an additive structure $\lambda(t \mid \bar{a}(t), z) = h_A(t, \bar{a}(t)) + h_Z(t, z)$ then it follows that
\begin{equation*}
    \begin{split}
        p_t(z \mid \bar{a}(t)) &= \frac{\exp\left(- \int_0^t h_Z(s, z) \d s \right) p(z)}{\int \exp\left(- \int_0^t h_Z(s, z) \d s \right) p(z) \d z} \\
        &= p_t(z),
    \end{split}
\end{equation*}
which does not depend on $\bar{a}(t)$, implying that $Z \ind \bar{A}(t) \mid T \geq t$. The marginal hazard is then given by
\begin{equation}
    \label{eq:hazard_marginal_additive}
    \begin{split}
        \lambda(t \mid \bar{a}(t)) &= E(\lambda(t \mid \bar{a}(t), Z) \mid \bar{A}(t) = \bar{a}(t), T \geq t) \\
            &= h_A(t, \bar{a}(t)) + E(h_Z(t, Z) \mid T \geq t),
    \end{split}
\end{equation}
so if the conditional hazard additionally has the Markov property, i.e. $h_A(t, \bar{a}(t)) = h_A(t, a(t-))$, then the marginal hazard has the Markov property as well! Since the discrete-time additive hazard model is not collapsible \citep{martinussenCollapsibilityConfoundingBias2013, sjolanderNoteNoncollapsibilityRate2016}, this construction only works in continuous time, and therefore does not provide a counter-example to the discrete-time argument of \citet{sjolanderCautionaryNoteExtended2020}.

As just demonstrated, the absence of treatment-frailty interaction on the hazard scale preserves the Markov property upon marginalization over $Z$. \Cref{result:no_effect_modification} shows that the converse, which is considerably less obvious, also holds: simultaneous validity of the conditional and marginal Markov properties necessarily imply this no-interaction condition, as long as the treatment process is properly time-dependent. The proof can be found in \Cref{sec:no_effect_modification_proof}.

\begin{theorem}
\label{result:no_effect_modification}
    Assume that $A(t)$ is a marked point process locally independent of $Z$ which satisfies \Cref{assumption:treatment_mixing} in \Cref{sec:no_effect_modification_proof}, and that $\lambda(t \mid \bar{a}(t), z) = h(t, a(t-), z)$ for some non-negative function $h(t, a, z)$. Then the following statements are equivalent:
    \begin{enumerate}
        \item $\lambda(t \mid \bar{a}(t)) = g(t, a(t-))$ for some non-negative function $g(t, a)$.
        \item $Z \ind \bar{A}(t) \mid T \geq t$.
        \item $h(t, a, z) = h_A(t, a) + h_Z(t, z)$ for some functions $h_A(t, a)$ and $h_Z(t, z)$.
    \end{enumerate}
\end{theorem}

\begin{remark}
     If the no-interaction condition is assumed along with conditional Markov then the marginal Markov property holds as stated in \Cref{result:no_effect_modification}. One might ask if a marginal proportional hazards model may also be correctly specified. This is possible, but it requires the conditional hazard to have a precise structure, which depends on the frailty distribution. Letting $\varphi_Z(r) = E(\e^{-Zr})$ denote the Laplace transform of the distribution of $Z$ and $c(t)$ any non-negative function with cumulative version $C(t) = \int_0^t c(s) \d s$, we show in \Cref{sec:marginal_Cox} that if the conditional hazard has the specific form
    \begin{equation*}
        \lambda(t \mid \bar{a}(t), z) = \lambda_0(t) \e^{\beta a(t-)} + z \frac{\d}{\d t} \varphi_Z^{-1}\left( \e^{- C(t)} \right) - c(t),
    \end{equation*}
    then the marginal hazard is $\lambda(t \mid \bar{a}(t)) = \lambda_0(t) \e^{\beta a(t-)}$, and thus the current-treatment Cox model is correctly specified. If $Z$ follows a Gamma distribution with mean 1 and variance $\theta$, the corresponding conditional hazard is
    \begin{equation*}
        \lambda(t \mid \bar{a}(t), z) = \lambda_0(t) \e^{\beta a(t-)} + z c(t) \e^{\theta C(t)} - c(t).
    \end{equation*}
    This construction shows that a marginal current-treatment Cox MSM can arise in the presence of unmeasured frailty. At the same time, it illustrates that it requires the conditional model to be calibrated to the frailty distribution. The same conditional event mechanism thus cannot induce a marginal current-treatment Cox model in a population with a different distribution of prognostic heterogeneity. Postulating the Cox MSM directly at the marginal level is therefore mathematically legitimate, but treats the proportional hazards property as population-specific structural assumption that cannot be justified from the conditional model alone.
\end{remark}

\begin{remark}
It is seen from \eqref{eq:hazard_marginal_additive} that, if there is no treatment-frailty interaction on the additive hazard scale, marginal Markov implies conditional Markov. It is possible, in principle, to have the Markov property hold marginally while the conditional hazard depends on past treatment, but only in the presence of treatment-frailty interaction on the hazard scale. In that case, marginal Markov would require the history-dependence of the conditional hazard to be exactly offset by history-dependent changes in the frailty distribution among survivors. Thus, dependence on past treatment may remain within strata defined by $Z$, but disappear after averaging over the frailty distribution among survivors. Because the required cancellation depends on the frailty distribution, it generally will not persist in another population with the same conditional event mechanism but a different distribution of prognostic heterogeneity. In other words, the only way to justify a population-agnostic marginal Markov property is to appeal to a conditional Markov property and have no effect modification on the hazard scale.
\end{remark}

If the Markov property is invoked both conditionally and marginally, \Cref{result:no_effect_modification} shows that no unmeasured frailty $Z$ can modify the treatment effect on the hazard scale. Since the marginal Markov property can only be etiologically justified by appealing to an underlying conditional Markov mechanism, a current-treatment Cox MSM implicitly rests on the absence of treatment-effect modification on the hazard scale by any unmeasured prognostic factor. This structural condition cannot be verified from the observed data and appears highly implausible in most applications. It resembles the no-effect-modification assumption needed in instrumental variable (IV) analyses to identify a population-average treatment effect, which rules out effect modification on the additive scale by unmeasured confounders \citep{hernanInstrumentsCausalInference2006}. However, we stress that $Z$ here is independent of the exposure, so it has nothing to do with unmeasured confounding. This arguably makes the requirement even stricter than in the IV setting.

\section{The rate function and extended Kaplan-Meier curves}
\label{sec:causal_rate}

We stress that the usual causal identification conditions require conditioning on the full treatment history, so the potential outcome hazard is identified via the observed hazard, and not the observed rate. Thus, while fitting a rate regression model avoids the Markov assumption altogether when studying associations, there is no such luxury when attempting causal inference. The rate, by definition, has no information about past treatment, and therefore it seems unlikely that it could in general be connected to a potential outcome under some treatment strategy. The rate function turns out to be connected to the extended Kaplan-Meier curve \citep{simonNonparametricGraphicalRepresentation1984, snapinnIllustratingImpactTimeVarying2005} which was proposed as a tool to visualize the effect of a time-varying covariate on survival, analogously to the way that ordinary Kaplan-Meier curves are used for baseline covariates. The idea is to let subjects swap risk sets over time according to the time-varying exposure $A(t)$, which is assumed to be discrete. More precisely, the extended Kaplan-Meier curves are defined for each exposure level $a$ as
\begin{equation*}
\label{eq:extended_KM}
    \hat{S}_a(t) = \prod_{u \in (0, t]} \left( 1 - \frac{\Delta N_{a}(u)}{Y_{a}(u)} \right),
\end{equation*}
where $\Delta N_a(t)$ is the number of events at time $t$ among people with $A(t) = a$, and $Y_a(t)$ is the number of people at risk at time $t$ with $A(t) = a$. The product is over the event times in $(0, t]$. In \Cref{sec:extended_KM_proof}, we show that the extended Kaplan-Meier curve estimates the exponential of minus the cumulative rate $R_a(t) = \int_0^t r(u \mid a) \d u$ for a fixed treatment level $a$, in the sense that
\begin{equation*}
    \hat{S}_a(t) \overset{P}{\to} \e^{-R_a(t)},
\end{equation*}
where $\overset{P}{\to}$ denotes convergence in probability. The interpretation of the extended Kaplan-Meier curve is therefore intimately tied to that of the rate function. As we have argued, the rate function does not have a direct causal interpretation without assuming the Markov property, and the same can thus be said for the extended Kaplan-Meier curve, confirming the claim made by \citet{sjolanderCautionaryNoteExtended2020}.

Under the Cox rate model \eqref{eq:cox_hazard_partly}, the ratio of the logarithm of the extended Kaplan-Meier curves for the two treatment levels therefore estimates the rate ratio
\begin{equation*}
    \frac{\log(\hat{S}_1(t))}{\log(\hat{S}_0(t))} \overset{P}{\to} \e^{\beta}.
\end{equation*}
Since the extended Kaplan-Meier curve does not have a causal interpretation without the Markov property, the same can be said for the parameter obtained from a Cox rate model. This remains true even if there is no confounding and the Cox rate model is correctly specified.

\section{Consequences of not capturing the effect of past treatment}
\label{sec:example}

To illustrate the importance of correctly modeling the dependence on the entire treatment history, and not just the current level, we present a hypothetical scenario where a proportional rates model \eqref{eq:cox_hazard_partly} holds, but where the hazard does not have the Markov property. The Cox rate model is therefore correctly specified, and common software for fitting Cox models can be used to do inference correctly. But the model is not equivalent to a Cox MSM with current treatment only, even in the absence of confounding, and the true average treatment effect does not equal what one would get from transforming the rate model to survival scale.

We consider a data generating process that can be described by an irreversible illness-death model. All subjects start without treatment (state 0) and can then start treatment at some (individual) random point in time (state 1), and once treatment is initiated, it cannot be discontinued. Death corresponds to state 2. The model is parametrized through the state transition hazards $\lambda_{ij}$ for $i, j \in \{0, 1, 2\}$ with $i < j$. As in \Cref{sec:markov_problem}, we also have an unmeasured frailty variable $Z$. The hazard of death is thus
\begin{equation*}
    \lambda(t \mid \bar{a}(t), z) = \begin{cases}
        \lambda_{02}(t \mid z) & \text{if } a(t-) = 0  \\
        \lambda_{12}(t \mid u, z) & \text{if } a(t-) = 1
    \end{cases}
\end{equation*}
where $u = \inf\{t \mid a(t) = 1\}$ is the time of treatment initiation. We assume that $Z$ affects the conditional outcome hazard multiplicatively, such that
\begin{equation*}
    \begin{split}
        &\lambda_{02}(t \mid z) = z h_0(t) \\
        &\lambda_{12}(t \mid u, z) = z h_1(t),
    \end{split}
\end{equation*}
for two non-negative functions $h_0(t), h_1(t)$. Note that the conditional hazard is Markov, since $\lambda_{12}(t \mid u, z)$ does not depend on $u$. As detailed in \Cref{sec:multiplicative_frailty}, the marginal (not conditional on $Z$) hazards $\lambda_{02}(t)$ and $\lambda_{12}(t \mid u)$ can be calculated from $h_0(t), h_1(t)$ and the Laplace transform of the frailty distribution, and the process will not be Markov when only conditioning on observed data. The corresponding observed rate among the untreated is $r(t \mid 0) = \lambda_{02}(t)$ (since there is only one possible treatment history among those with $A(t) = 0$; they were always untreated). The rate among the treated can be found with \eqref{eq:rate_from_hazard} as $r(t \mid 1) = E(\lambda_{12}(t \mid U) \mid T \geq t, U < t)$. Finally, we assume that $\lambda_{01}(t \mid z) = \lambda_{01}(t)$ does not depend on $z$, in order for treatment to be locally independent of $Z$.

The model is completely specified by $h_0(t), h_1(t), \lambda_{01}(t)$ and the frailty distribution. The question is whether one can choose these such that the resulting observed rates solve the proportional rates equation
\begin{equation}
\label{eq:prop_rate}
    r(t \mid 1) = \e^{\beta} r(t \mid 0)
\end{equation}
for some choice of $\beta$. With $h_1(t), \lambda_{01}(t), \beta$ and frailty distribution fixed, one could in principle try to solve \eqref{eq:prop_rate} explicitly for $h_0(t)$, but this is not trivial. Instead, a numerical procedure can be employed that iteratively updates $h_0(t)$ in a way such that \eqref{eq:prop_rate} is fulfilled. We describe this procedure in \Cref{sec:cox_rate_construction}.

\begin{figure}
    \centering
    \includegraphics[width=\linewidth]{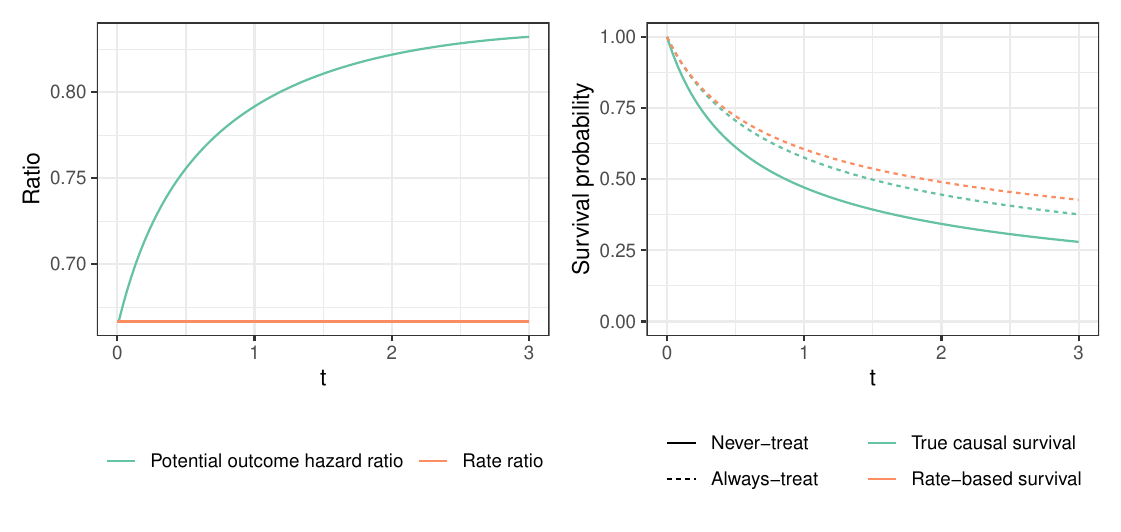}
    \caption{Left: The rate ratio that would be estimated by a Cox MSM vs. the true causal hazard ratio comparing the always-treat and never-treat strategies. Right: The survival probability estimated by transforming the rate estimated by a Cox MSM to survival scale vs. the true survival probability under the always-treat and never-treat strategies. The curves are identical for the never-treat strategy.}
    \label{fig:prop_rate_example}
\end{figure}

For the sake of illustration, we choose a Gamma frailty distribution with mean $1$ and variance $2$, $\lambda_{01}(t) = 0.3$, $h_1(t) = 1$ and $\beta = \log \left( \frac{2}{3} \right)$. \Cref{fig:prop_rate_example}, left panel, shows the resulting rate ratio, constantly equal to $2/3$ by design, and the ratio of the potential outcome hazards under the always-treat and never-treat regimes, that is the ratio $\lambda_{12}(t \mid 0)/\lambda_{02}(t)$, which is strictly larger than $2/3$ for $t > 0$. \Cref{fig:prop_rate_example}, right panel, shows the potential outcome survival function under both treatment strategies, and the survival function that would be obtained by transforming the rate to survival scale through taking the exponential of minus the cumulative rate (this is what would be estimated by the extended Kaplan-Meier curve in this setting). For the never-treat strategy, the two curves are identical as the rate equals the hazard. However, under the always-treat strategy, the two curves are different. At $t = 3$ the correct causal contrast is $P(T^{\bar{1}} > t) - P(T^{\bar{0}} > t) = 0.10$, while the difference of the rate-based curves is $0.15$. This highlights the fact that although the dependence on current treatment is correctly specified, the model does not lead to correct inference about causal quantities since the dependence on the treatment history is not correctly specified.

\section{Discussion}
\label{sec:discussion}

In this paper, we have revisited Sj{\"o}lander's warning concerning the causal validity of proposed extensions of the Kaplan-Meier estimator to time-varying exposures. We have broadened this warning to encompass Cox regression with time-varying treatment and current-treatment Cox MSMs, and sharpened it by characterizing the conditions under which the Markov property can hold in continuous time. We show that the conditional and marginal Markov properties can hold simultaneously only in the absence of treatment-frailty interaction on the hazard scale. Thus, unlike in the discrete-time setting considered by \citet{sjolanderCautionaryNoteExtended2020}, a treatment effect and non-degenerate frailty may coexist with the Markov property, but only under an additive no-interaction condition involving unmeasured factors. This requirement is necessary in order for the Markov property to not be a population-specific assumption.

We stress that our result has direct implications for the widely used Cox MSMs that include only current treatment. Although the weighting used in an MSM may appropriately address measured time-dependent confounding, it does not remove the need for the Markov property when the fitted model is used to derive intervention-specific survival probabilities. A causal interpretation of the resulting survival probabilities therefore rests on the absence of treatment-effect modification on the additive hazard scale by any unmeasured prognostic factor. This structural condition cannot be verified from the observed data and appears fundamentally implausible in realistic applications. Moreover, we have demonstrated that a current-treatment Cox model may be correctly specified as a model for the observed-data rate even when the Markov property fails. Correct specification of this rate model therefore does not ensure that survival probabilities derived from it correspond to the true intervention-specific survival probabilities.

The issue is not caused by the proportional hazards structure of the Cox model. It arises whenever the intensity conditional on the entire treatment history is replaced by a model involving only current treatment, thereby assuming the Markov property. Similar considerations therefore apply to multi-state models with an absorbing state and to recurrent-event models with a terminal event. Such models are increasingly important in biomedical research; for example, the European Medicines Agency has emphasized the relevance of treatment effects on both first and subsequent events when advocating recurrent-event endpoints \citep{akachaRequestCHMPQualification2018}. We detail the corresponding implications for additive hazard models in \Cref{sec:Additive rates}.

The multi-state model literature has for a while acknowledged the fact that many methods rely on the Markov property and that it might not be satisfied. This had lead to work on non-Markov methods for estimating state transition probabilities \citep{andersenInferenceTransitionProbabilities2022, deuna-alvarezNonparametricEstimationTransition2015, meira-machadoNonparametricEstimationTransition2006, putterNonparametricEstimationTransition2018, titmanTransitionProbabilityEstimates2015}. More recently, there has been some work on testing the (marginal) Markov property \citep{rodriguez-girondoNonparametricTestMarkovianity2012, soutinhoMethodsCheckingMarkov2022, titmanGeneralTestsMarkov2022, soutinhoMarkovMSMPackageChecking2023}. This seems motivated by a sentiment that ``in the multi-state setting, it is of practical interest to determine whether the Markov property holds within a particular data set'' \citep{soutinhoMarkovMSMPackageChecking2023}. This line of thinking treats the marginal Markov property as an attribute specific to each dataset, rather than an emergent feature of the data-generating process. A conditional Markov property is more substantive since it allows a marginal version to hold in more than a single population, but that requires the assumption of no treatment-frailty interaction, which is not testable in any reasonable experiment.

\section*{Funding}

MBK is funded by Novo Nordisk Fonden grant NNF22OC0076595.

\printbibliography

\pagebreak
\appendix

\section{Results on continuous-time hazards}
\label{sec:marginal_hazards_frailty}

We here give some results on the relationship between the full- and marginal hazard functions. We consider an exposure process $A(t)$, a time-to-event outcome $T$ with associated counting process $N(t)$, and an unobserved non-negative frailty variable $Z$. We assume that $A(t)$ is a marked point process, i.e. that it is piecewise constant with values updated every time the associated counting process $N^A(t)$ jumps. The distribution of the times of exposure updates is assumed to be absolutely continuous with hazard $\gamma^A(t \mid \bar{a}(t), z)$ conditional on $T \geq t$, $\bar{A}(t) = \bar{a}(t)$ and $Z = z$. The distribution of the updated exposure value $A(t)$ conditional on $T \geq t$, $\Delta N^A(t) = 1$, $\bar{A}(t) = \bar{a}(t)$ and $Z = z$ is denoted $\mu^A_t(a \mid \bar{a}(t), z)$. We define $\gamma^A(t \mid \bar{a}(t))$ and $\mu^A_t(a \mid \bar{a}(t))$ analogously, but not conditional on $Z$, and denote by $\Gamma^A(t \mid \bar{a}(t), z)$ and $\Gamma^A(t \mid \bar{a}(t))$ the cumulative hazards. 

\subsection{Local independence}
\label{sec:local_independence}

The following local independence property \citep{didelezGraphicalModelsMarked2008} captures the assumption that frailty and exposure are independent.

\begin{definition}[Local independence]
\label{def:local_independence}
    We say that $A(t)$ is locally independent of $Z$ if $\gamma^A(t \mid \bar{a}(t), z) = \gamma^A(t \mid \bar{a}(t))$ and $\mu^A_t(a \mid \bar{a}(t), z) = \mu^A_t(a \mid \bar{a}(t))$. 
\end{definition}

\subsection{Distribution of frailty among survivors}
\label{sec:frailty_survivor}

\begin{lemma}
\label{result:Z_distribution_survivor}
    Assume that $A(t)$ is a marked point process locally independent of $Z$. Then the conditional distribution of $Z \mid T \geq t, \bar{A}(t) = \bar{a}(t)$ has density
    \begin{equation}
    \label{eq:Z_density_survivor}
        \begin{split}
            p_t(z \mid \bar{a}(t)) &= \exp\left( \Lambda(t \mid \bar{a}(t)) - \Lambda(t \mid \bar{a}(t), z) \right) p(z) \\
            &= \frac{\exp\left( - \Lambda(t \mid \bar{a}(t), z) \right) p(z)}{\int \exp\left( - \Lambda(t \mid \bar{a}(t), z) \right) p(z) \d z},
        \end{split}
    \end{equation}
    and the marginal hazard at time $t$ is given by
    \begin{equation}
        \begin{split}
            \lambda(t \mid \bar{a}(t)) &= \int \lambda(t \mid \bar{a}(t), z) p_t(z \mid \bar{a}(t)) \d z \\
            &= \frac{\int \lambda(t \mid \bar{a}(t), z)\exp\left(- \Lambda(t \mid \bar{a}(t), z) \right) p(z) \d z}{\int \exp\left(- \Lambda(t \mid \bar{a}(t), z) \right) p(z) \d z}.
        \end{split}
    \end{equation}
\end{lemma}
\begin{proof}
    The likelihood of a realization $(T \geq t, \bar{A}(t), Z)$ is \citep[chapter II.7 \& IX.3]{andersenStatisticalModelsBased1993}
    \begin{equation*}
        \begin{split}
        	\d P(T \geq t, \bar{A}(t), Z) &= p(Z) \exp(- \Lambda(t \mid \bar{A}(t), Z)) \\
            &\qquad \exp(- \Gamma^A(t \mid \bar{A}(t), Z)) \prod_{s < t} \left( \gamma^A(s \mid \bar{A}(s), Z) \d s \d \mu^A_t(A(s) \mid \bar{A}(s), Z) \right)^{\Delta N^A(s)} \\
            &= p(Z) \exp(- \Lambda(t \mid \bar{A}(t), Z)) \\
            &\qquad \exp(- \Gamma^A(t \mid \bar{A}(t))) \prod_{s < t} \left( \gamma^A(s \mid \bar{A}(s)) \d s \d \mu^A_t(A(s) \mid \bar{A}(s)) \right)^{\Delta N^A(s)},
        \end{split}
    \end{equation*}
    where $p(z)$ is the density of $Z$ and $\Delta N^A(t) = N^A(t) - N^A(t-)$ the increment of $N^A(t)$. Likewise, the marginal likelihood of $(T \geq t, \bar{A}(t))$ is
    \begin{equation*}
    	\begin{split}
    	    \d P(T \geq t, \bar{A}(t)) &= \exp(- \Lambda(t \mid \bar{A}(t))) \exp(- \Gamma^A(t \mid \bar{A}(t))) \prod_{s < t} \left( \gamma^A(s \mid \bar{A}(s)) \d s \d \mu^A_t(A(s) \mid \bar{A}(s)) \right)^{\Delta N^A(s)}.
    	\end{split}
    \end{equation*}
    The conditional distribution of $Z$ given $T \geq t, \bar{A}(t) = \bar{a}(t)$ therefore has density
    \begin{equation*}
    	\begin{split}
    	    p_t(z \mid \bar{a}(t)) &= \frac{\d P(T \geq t, \bar{a}(t), z)}{\d P(T \geq t, \bar{a}(t))} \\
            &= \exp(\Lambda(t \mid \bar{a}(t)) - \Lambda(t \mid \bar{a}(t), z)) p(z).
    	\end{split}
    \end{equation*}
    Integrating both sides over $z$ and rearranging yields
    \begin{equation*}
        \exp(- \Lambda(t \mid \bar{a}(t))) = \int \exp(- \Lambda(t \mid \bar{a}(t), z)) p(z) \d z,
    \end{equation*}
    which shows \eqref{eq:Z_density_survivor}. 
    
    The hazard $\lambda(t \mid \bar{a}(t))$ can be found via an iterated expectation (or more formally, the innovation theorem \citep[p. 80]{andersenStatisticalModelsBased1993}) as
    \begin{equation*}
    	\begin{split}
    	    \lambda(t \mid \bar{a}(t)) &= E(\lambda(t \mid \bar{a}(t), Z) \mid \bar{A}(t) = \bar{a}(t), T \geq t) \\
            &= \int \lambda(t \mid \bar{a}(t), z) p_t(z \mid \bar{a}(t)) \d z \\
            &= \frac{\int \lambda(t \mid \bar{a}(t), z) \exp(- \Lambda(t \mid \bar{a}(t), z)) p(z) \d z}{\int \exp(- \Lambda(t \mid \bar{a}(t), z)) p(z) \d z},
    	\end{split}
    \end{equation*}
    where we use the result above.
\end{proof}

\subsection{Proof of \Cref{result:no_effect_modification}}
\label{sec:no_effect_modification_proof}

The following assumption ensures that the treatment $A(t)$ is properly time-dependent, by requiring that the hazard of a treatment change is always non-zero until an absorbing state is hit. This rules out the baseline treatment setting $A(t) = A$ where both the conditional and marginal Markov property are automatically satisfied, but \Cref{result:no_effect_modification} obviously does not always hold. Further, we require that the set of possible treatment values is ``properly connected''.

\begin{assumption}
\label{assumption:treatment_mixing}
    Assume that $\gamma^A(t \mid \bar{a}(t)) > 0$ for all $t$ except if $a(t-)$ is an absorbing state for the treatment, and that the support of $\mu_t^A(a \mid \bar{a}(t))$ depends only on $a(t-)$. Further, assume that the set of possible treatment values is connected in the sense that for any two values $a$, $a'$, there exists a sequence $\{a_k\}_{k=0}^K$ such that $a_0 = a$, $a_K = a'$ and for all $k \in \{0, \dots, K-1\}$ at least one of the treatment changes $a_k \to a_{k+1}$ or $a_{k+1} \to a_k$ are possible.
\end{assumption}

\begin{proof}[Proof of \Cref{result:no_effect_modification}]
    As in \Cref{result:Z_distribution_survivor} in \Cref{sec:frailty_survivor} we denote the density of $Z \mid T \geq t, \bar{A}(t) = \bar{a}(t)$ by $p_t(z \mid \bar{a}(t))$.
    
    $1 \implies 2$: Assume that $\lambda(t \mid \bar{a}(t)) = g(t, a(t-))$, take any two valid histories $\bar{a}(t)$ and $\bar{a}'(t)$ and let $a = a(t-)$ and $a' = a'(t-)$. 
    
    Consider first the case where $a = a'$. Then
    \begin{equation*}
        \begin{split}
            0 &= g(t, a) - g(t, a') \\
            &= \lambda(t \mid \bar{a}(t)) - \lambda(t \mid \bar{a}'(t)) \\
            &= \int h(t, a, z) (p_t(z \mid \bar{a}(t)) - p_t(z \mid \bar{a}'(t))) \d z,
        \end{split}
    \end{equation*}
    where we used \Cref{result:Z_distribution_survivor}. Since $p_t(z \mid \bar{a}(t))$ and $p_t(z \mid \bar{a}'(t))$ do not depend on the value of $h(t, a, z)$ at time $t$, we can let it vary over all non-negative functions $f(z)$ for which the integral exists, and in all cases it must be true that
    \begin{equation*}
        0 = \int f(z) (p_t(z \mid \bar{a}(t)) - p_t(z \mid \bar{a}'(t))) \d z.
    \end{equation*}
    We conclude that $p_t(z \mid \bar{a}(t)) = p_t(z \mid \bar{a}'(t))$.

    Consider now the case where $a \neq a'$, but the treatment change $a \to a'$ is possible. Choose some $\varepsilon > 0$ small enough that $a(s) = a$ for all $s \in (t - 2\varepsilon, t)$ and define the history $\bar{a}_\varepsilon(t)$ by $a_\varepsilon(s) = a(s)$ for $s < t - \varepsilon$ and $a_\varepsilon(s) = a'$ for $s \in [t - \varepsilon, t)$. The history $\bar{a}_\varepsilon(t)$ is valid as the jump $a \to a'$ is possible. Since $a'(t-) = a' = a_\varepsilon(t-)$ we get from the result above that $p_t(z \mid \bar{a}'(t)) = p_t(z \mid \bar{a}_\varepsilon(t))$. We then get that
    \begin{equation*}
        \begin{split}
            \frac{p_t(z \mid \bar{a}(t))}{p_t(z \mid \bar{a}'(t))} &= \frac{p_t(z \mid \bar{a}(t))}{p_t(z \mid \bar{a}_\varepsilon(t))} \\
            &= \exp\left( \int_{t - \varepsilon}^t g(s, a) - h(s, a, z) - (g(s, a') - h(s, a', z)) \d s \right),
        \end{split}
    \end{equation*}
    where the second equality uses \Cref{result:Z_distribution_survivor} and the fact that $\bar{a}(t)$ and $\bar{a}_\varepsilon(t)$ agree up to time $t - \varepsilon$. Taking the limit $\varepsilon \to 0$ we see that $p_t(z \mid \bar{a}(t)) = p_t(z \mid \bar{a}'(t))$.

    Finally, consider the general case where $a \neq a'$. By \Cref{assumption:treatment_mixing}, there must be some sequence $\{a_k\}_{k=0}^K$ such that $a_0 = a$ and $a_K = a'$ and for all $k \in \{0, \dots, K-1\}$ at least one of the changes $a_k \to a_{k+1}$ or $a_{k+1} \to a_k$ are possible. For each $k \in \{1, \dots, K-1\}$ let $\bar{a}_k(t)$ be any history for which $a_k(t-) = a_k$. The result above implies that $p_t(z \mid \bar{a}(t)) = p_t(z \mid \bar{a}_1(t))$, that $p_t(z \mid \bar{a}_k(t)) = p_t(z \mid \bar{a}_{k+1}(t))$ for $k \in \{1, \dots, K-2\}$ and that $p_t(z \mid \bar{a}_{K-1}(t)) = p_t(z \mid \bar{a}'(t))$. This shows that $p_t(z \mid \bar{a}(t)) = p_t(z \mid \bar{a}'(t))$.

    $2 \implies 1$: Assume that $Z \ind \bar{A}(t) \mid T \geq t$ such that $p_t(z \mid \bar{a}(t)) = p_t(z)$. Then
    \begin{equation*}
        \begin{split}
            \lambda(t \mid \bar{a}(t)) &= \int h(t, a(t-), z) p_t(z) \d z \\
            &= g(t, a(t-)),
        \end{split}
    \end{equation*}
    using \Cref{result:Z_distribution_survivor}.

    $1 \implies 3$: Assume that $\lambda(t \mid \bar{a}(t)) = g(t, a(t-))$ which we have now established is equivalent to assuming $p_t(z \mid \bar{a}(t)) = p_t(z)$. From \Cref{result:Z_distribution_survivor} we get that
    \begin{equation*}
        \begin{split}
            \frac{\d}{\d t} p_t(z) &= \frac{\d}{\d t} p_t(z \mid \bar{a}(t)) \\
            &= (\lambda(t \mid \bar{a}(t)) - \lambda(t \mid \bar{a}(t), z))  \exp\left( \Lambda(t \mid \bar{a}(t)) - \Lambda(t \mid \bar{a}(t), z) \right) p(z)  \\
            &= (g(t, a(t-)) - h(t, a(t-), z)) p_t(z).
        \end{split}
    \end{equation*}
    Since $\frac{\d}{\d t} p_t(z)$ and $p_t(z)$ do not depend on $a(t-)$, neither can $g(t, a(t-)) - h(t, a(t-), z)$, implying that $g(t, a(t-)) - h(t, a(t-), z) = -h_Z(t, z)$ for some function $h_Z(t, z)$, or equivalently $h(t, a(t-), z) = g(t, a(t-)) + h_Z(t, z)$.

    $3 \implies 2$: Assume that $h(t, a, z) = h_A(t, a) + h_Z(t, z)$. Then it follows from \Cref{result:Z_distribution_survivor} that
    \begin{equation*}
        \begin{split}
            p_t(z \mid \bar{a}(t)) &= \frac{\exp\left(- \int_0^t h_Z(s, z) \d s \right) p(z)}{\int \exp\left(- \int_0^t h_Z(s, z) \d s \right) p(z) \d z} \\
            &= p_t(z),
        \end{split}
    \end{equation*}
    such that $Z \ind \bar{A}(t) \mid T \geq t$.
\end{proof}

\subsection{The multiplicative case}
\label{sec:multiplicative_frailty}

Assume now that the $Z$-conditional hazard is Markov with a multiplicative effect of $Z$, such that
\begin{equation}
\label{eq:hazard_multiplicative_frailty}
    \lambda(t \mid \bar{a}(t), z) = z h(t, a(t-)),
\end{equation}
where $h(t, a)$ is some non-negative function with cumulative version $H(t, \bar{a}(t)) = \int_0^t h(s, a(s-)) \d s$. Since $Z$ modifies the effect of $A(t)$ on the hazard, we know from \Cref{result:no_effect_modification} that the marginal hazard cannot be Markov except if $Z$ has a degenerate distribution or there is no effect of treatment. 

Note that this conclusion relies crucially on the assumption that $\lambda(t \mid \bar{a}(t), z)$ is Markov, i.e. that $h(t, a(t-))$ does not depend on $\bar{a}(t)$. It is possible to have a hazard $\lambda(t \mid \bar{a}(t))$ which is Markov by choosing a history-dependent $\tilde{h}(t, \bar{a}(t))$ as follows. Using \Cref{result:Z_distribution_survivor} with \eqref{eq:hazard_multiplicative_frailty} we see that
\begin{equation}
\label{eq:marginal_hazard_Laplace}
    \begin{split}
    \lambda(t \mid \bar{a}(t)) &= -\frac{\d}{\d t} \log\left( \varphi_Z(\tilde{H}(t, \bar{a}(t))) \right),
    \end{split}
\end{equation}
where $\varphi_Z(r) = E(\e^{-Zr})$ is the Laplace-transform of the distribution of $Z$. We can then obtain a marginal Markov hazard $\lambda(t \mid \bar{a}(t)) = g(t, a(t-))$ by choosing the conditional hazard
\begin{equation*}
    \tilde{h}(t, \bar{a}(t)) = \frac{\d}{\d t} \varphi_Z^{-1}\left( \exp \left( -\int_0^t g(s, a(s-)) \d s \right) \right),
\end{equation*}
which is obtained by solving \eqref{eq:marginal_hazard_Laplace} for $\tilde{h}$.

\subsection{Marginal Cox model from conditional additive model}
\label{sec:marginal_Cox}

We here show how to construct a marginal Markov Cox model from a conditional Markov hazard with no treatment-frailty interaction. If we take 
\begin{equation*}
    \lambda(t \mid \bar{a}(t), z) = \lambda_0(t) \e^{\beta a(t-)} + h_Z(t, z),
\end{equation*}
then by \Cref{result:no_effect_modification} the resulting marginal hazard will be Markov because of the additive structure of the conditional hazard function. As in \eqref{eq:hazard_marginal_additive}, the marginal hazard is
\begin{equation*}
    \lambda(t \mid \bar{a}(t)) = \lambda_0(t) \e^{\beta a(t-)} + E(h_Z(t, z) \mid T \geq t),
\end{equation*}
so if the function $h_Z$ is calibrated in such a way that $E(h_Z(t, z) \mid T \geq t) = 0$ then the conditional hazard marginalizes to a current-treatment Cox model. Consider for example the case where $h_Z(t, z) = z b(t) - c(t)$ for some functions $b(t)$ and $c(t)$. The required restriction is equivalent to $b(t)$ satisfying the equation
\begin{equation*}
    \begin{split}
        c(t) &= b(t) E(Z \mid T \geq t) \\
        &= - \frac{\d}{\d t} \log\left( \varphi_Z\left( B(t) \right) \right),
    \end{split}
\end{equation*}
where $B(t) = \int_0^t b(s) \d s$ and $\varphi_Z(r) = E(\e^{-Z r})$ is the Laplace transform of the distribution of $Z$. The solution is
\begin{equation*}
    b(t) = \frac{\d}{\d t} \varphi_Z^{-1}\left( \e^{- C(t)} \right),
\end{equation*}
where $C(t) = \int_0^t c(s) \d s$. This implies that
\begin{equation*}
    \lambda(t \mid \bar{a}(t), z) = \lambda_0(t) \e^{\beta a(t-)} + z \frac{\d}{\d t} \varphi_Z^{-1}\left( \e^{- C(t)} \right) - c(t)
\end{equation*}
marginalizes to $\lambda(t \mid \bar{a}(t)) = \lambda_0(t) \e^{\beta a(t-)}$ for any choice of $c(t)$ and frailty distribution with corresponding Laplace transform $\varphi_Z$. For instance, if $Z$ has a Gamma distribution with mean 1 and variance $\theta$ then the conditional hazard should be
\begin{equation*}
    \lambda(t \mid \bar{a}(t), z) = \lambda_0(t) \e^{\beta a(t-)} + z c(t) \e^{\theta C(t)} - c(t),
\end{equation*}
in order to give a marginal current-treatment Cox model.

\section{The extended Kaplan-Meier curve and the rate ratio}
\label{sec:extended_KM_proof}

We here show that the extended Kaplan-Meier curve estimates the exponential of the negative cumulative rate function. Each observation $(T_i, \Delta_i, \{A_i(t)\}_{t \geq 0})$, where $\Delta_i = I(T^*_i \leq C_i)$ is an indicator of the event $T^*_i$ happening before the censoring time $C_i$, can be represented in counting process format as $\{(Y_i(t), N_i(t), A_i(t))\}_{t \geq 0}$, where $Y_i(t) = I(t \leq T_i)$ is a predictable at-risk indicator and $N_i(t) = I(T_i \leq t, \Delta_i = 1)$ indicates an event having occurred at- or before time $t$. We assume an iid. sample of $n$ such observations.

Define for each treatment level $a$ the derived processes $N_{a,i}(t) = I(T_i \leq t, \Delta_i = 1, A_i(T_i) = a) = N_i(t) I(A_i(T_i) = a)$ that count event occurrences stratified according to treatment status at the event time. Let also $Y_{a,i}(t) = I(t \leq T_i, A_i(t) = a) = Y_i(t) I(A_i(t) = a)$ be the treatment-specific at-risk indicators. Note that $N_i(t) = \sum_a N_{a,i}(t)$ and $Y_i(t) = \sum_a Y_{a,i}(t)$. Define the aggregate versions $N_a(t) = \sum_i N_{a,i}(t)$ and $Y_a(t) = \sum_i Y_{a,i}(t)$.

Let the treatment-specific Nelson-Aalen estimator be given as
\begin{equation}
	\hat{R}_a(t) =   \int_0^t \frac{J_a(u)}{Y_{a}(u)} \d  N_{a}(u),
\end{equation}
where $J_a(t) = I(Y_{a}(t) > 0)$. The extended Kaplan-Meier estimator can then be written as a product integral \parencite{gillSurveyProductIntegrationView1990} of the Nelson-Aalen estimator
\begin{equation*}
	\hat{S}_a(t) = \prodi_{u \in [0, t]} (1 - \d \hat{R}_a(u)).
\end{equation*}
We assume the following regularity condition.
\begin{assumption}
\label{assumption:EKM_regularity}
    We assume that there exists an integrable function $k$ that dominates the hazard on $[0, t]$ in the sense that $\lambda(t \mid \bar{A}(u)) \leq k(u)$ for $u \leq t$ with probability 1 and $\int_0^t k(u) \d u < \infty$. Assume for each treatment level $a$ also the following regularity conditions:
	\begin{equation*}
		\int_0^t \frac{J_a(u)}{Y_{a}(u)} k(u) \d u \overset{P}{\to} 0, \quad \int_0^t (1 - J_a(u)) k(u) \d u \overset{P}{\to} 0,
	\end{equation*}
	where $\overset{P}{\to}$ denotes convergence in probability as $n \to \infty$. 
\end{assumption}
We then have the following result.
\begin{lemma}
	\label{thm:Nelson-Aalen}
	Under \Cref{assumption:EKM_regularity} it holds that
	\begin{equation*}
		\sup_{u \in [0, t]} \left\lvert \hat{R}_a(u) - R_a(u) \right\rvert \overset{P}{\to} 0.
	\end{equation*}
\end{lemma}
\begin{proof}
	The time-to-event setup with a discrete time-varying treatment $A(t) \in \{0, \dots, k-1\}$ corresponds to a multi-state model in the following sense: let $X(t)$ be a multi-state process with $k + 1$ states:
	\begin{equation*}
		X(t) = \begin{cases}
			a & \text{if } A(t) = a \text{ and } N(t) = 0 \\
			k & \text{if } N(t) = 1
		\end{cases}
	\end{equation*}
	With this notation, the theorem is exactly that of \textcite[theorem 2.1]{niesslStatisticalInferenceState2023} for what they refer to as the cumulative partly conditional transition rates for the state-transition $a \to k$.
\end{proof}

With this result in hand, it follows from properties of product integration that the extended Kaplan-Meier estimator converges to the exponential of the cumulative rate.

\begin{proposition}
	Under \Cref{assumption:EKM_regularity} it holds that
	\begin{equation*}
		\hat{S}_a(t) \overset{P}{\to} \e^{-R_a(t)}.
	\end{equation*}
\end{proposition}

\begin{proof}
	Since the product integral is a continuous operator in supremum norm \citep[theorem 7]{gillSurveyProductIntegrationView1990}, the conclusion of \Cref{thm:Nelson-Aalen} implies that
	\begin{equation*}
		\hat{S}_a(t) = \prodi_{u \in [0, t]} (1 - \d \hat{R}_a(u)) \overset{P}{\to} \prodi_{u \in [0, t]} (1 - \d R_a(u)) = \e^{-R_a(t)}.
	\end{equation*}
\end{proof}

\section{Constructing a non-Markov proportional rates model}
\label{sec:cox_rate_construction}

One might wonder if there even exists joint distributions of $(\bar{A}(t), \bar{N}(t))$ with the proportional rates property
\begin{equation*}
	r(t \mid a(t-)) = r_0(t)  \e^{\beta a(t-)}.
\end{equation*}
The answer is trivially yes, since a Markovian proportional hazards model, which it is not hard to construct, implies a proportional rates model. A more interesting question is whether it can be done without the Markov property. We discuss the difficulties in constructing such an example explicitly, and then give a numerical procedure which seems to work. We then extend the numerical procedure to the case where the hazards are generated by unmeasured frailty.

We first rephrase the problem as an illness-death model with states $\{0,1,2\}$ corresponding to untreated, treated and dead, and corresponding state transition hazards $\lambda_{01}(t), \lambda_{02}(t)$ and $\lambda_{12}(t \mid U)$ where $U$ is the time of arrival into state 1. The process $X(t)$ indicates the current state. Note that the hazards going out of state 0 cannot depend on the past, and therefore in particular $r_{02}(t) = \lambda_{02}(t)$ is the rate of death for untreated.

We can calculate rate of death among the treated as follows:
\begin{equation}
	\label{eq:rate_calc}
	\begin{split}
		r_{12}(t) &= E(\lambda_{12}(t \mid U) \mid X(t) = 1) \\
		&= \int \lambda_{12}(t \mid u) p(U = u \mid X(t) = 1) \d u \\
		&= \int \lambda_{12}(t \mid u) \frac{p(U = u, X(t) = 1)}{P(X(t) = 1)} \d u \\
		&= \frac{\int_{0}^{t} \lambda_{12}(t \mid u) p(U = u, T \geq t) \d u}{P_{01}(0, t)} \\
		&= \frac{\int_{0}^{t} P_{00}(0, u) P_{11}(u, t \mid u) \lambda_{12}(t \mid u) \d \Lambda_{01}(u)}{P_{01}(0, t)} \\
		&= \frac{\int_{0}^{t} \mathrm{e}^{-(\Lambda_{01}(u) + \Lambda_{02}(u))} \mathrm{e}^{-\int_{u}^{t} \lambda_{12}(s \mid u) \d s } \lambda_{12}(t \mid u) \d \Lambda_{01}(u)}{\int_{0}^{t} \mathrm{e}^{-(\Lambda_{01}(u) + \Lambda_{02}(u))} \mathrm{e}^{-\int_{u}^{t} \lambda_{12}(s \mid u) \d s } \d \Lambda_{01}(u)}.
	\end{split}
\end{equation}
Here $P_{lm}(s, t) = P(X(t) = m \mid X(s) = l)$ are the state transition probabilities, with $P_{11}(s, t \mid u) = P(X(t) = 1 \mid X(s) = 1, U = u)$ depending also on the time of arrival to state 1.

\subsection{An integral equation}

A proportional rates model requires that $r_{12}(t) = \lambda_{02}(t) \mathrm{e}^{\beta}$ for some $\beta$, i.e. that
\begin{equation*}
	\begin{split}
		\int_{0}^{t} P_{00}(0, u) P_{11}(u, t \mid u) \lambda_{12}(t \mid u) \d \Lambda_{01}(u) &= \mathrm{e}^{\beta} \lambda_{02}(t) P_{01}(0, t) \\
		&= \mathrm{e}^{\beta} \lambda_{02}(t) \int_{0}^{t} P_{00}(0, u) P_{11}(u, t \mid u) \d \Lambda_{01}(u).
	\end{split}
\end{equation*}
We note that the LHS can be rewritten using Leibniz' rule as
\begin{equation*}
	\begin{split}
		\int_{0}^{t} & P_{00}(0, u) P_{11}(u, t \mid u) \lambda_{12}(t \mid u) \d \Lambda_{01}(u) \\
		&= -\int_{0}^{t} P_{00}(0, u) \frac{\d}{\d t} P_{11}(u, t \mid u) \d \Lambda_{01}(u) \\
		&= -\frac{\d}{\d t} \int_{0}^{t} P_{00}(0, u) P_{11}(u, t \mid u) \d \Lambda_{01}(u) + P_{00}(0, t) \lambda_{01}(t) P_{11}(t, t \mid t) \\
		&= -\frac{\d}{\d t} P_{01}(0, t) + P_{00}(0, t) \lambda_{01}(t).
	\end{split}
\end{equation*}
The equation we want to solve can thus be written as
\begin{equation*}
	\frac{\d}{\d t} P_{01}(0, t) = -\mathrm{e}^{\beta} \lambda_{02}(t) P_{01}(0, t) + P_{00}(0, t) \lambda_{01}(t),
\end{equation*}
which is an inhomogeneous linear ODE of order 1 with the solution
\begin{equation*}
	P_{01}(0, t) = \mathrm{e}^{-\Lambda_{02}(t) \mathrm{e}^{\beta}} \left( c + \int_{0}^{t} P_{00}(0, s) \lambda_{01}(s) \mathrm{e}^{\Lambda_{02}(s) \mathrm{e}^{\beta}} \d s \right) 
\end{equation*}
for some $c$. Using the fact that $P_{01}(0, 0) = 0$ we see that $c = 0$ such that
\begin{equation*}
	P_{01}(0, t) = \mathrm{e}^{-\Lambda_{02}(t) \mathrm{e}^{\beta}} \int_{0}^{t} P_{00}(0, s) \lambda_{01}(s) \mathrm{e}^{\Lambda_{02}(s) \mathrm{e}^{\beta}} \d s.
\end{equation*}
Replacing $P_{01}(0, t)$ with its definition we get the equation
\begin{equation*}
	\int_{0}^{t} P_{00}(0, u) P_{11}(u, t \mid u) \d \Lambda_{01}(u) =  \int_{0}^{t} \mathrm{e}^{-\mathrm{e}^{\beta} \int_ u^t \lambda_{02}(s) \d s} P_{00}(0, u) \d \Lambda_{01}(u),
\end{equation*}
which seems simpler than the original one, since $P_{11}(u, t \mid u)$ only enters into one side. However, it is still not obvious how to find a solution $(\lambda_{01}, \lambda_{02}, \lambda_{12})$ to the equation, except for the Markov case where $\lambda_{12}(t \mid u) = \lambda_{12}(t)$ does not depend on $u$.

\subsection{A numerical solution}
\label{sec:numerical_rates}

The problem is completely specified by a choice of $\beta$ and the transition hazards $\lambda_{01}(t), \lambda_{02}(t)$ and $\lambda_{12}(t \mid U)$. One can try to find a solution numerically by fixing e.g. $\lambda_{01}(t), \lambda_{12}(t \mid U)$ and then iteratively updating $\lambda_{02}(t)$ such that \eqref{eq:cox_hazard_partly} holds. This inspires the following procedure:
\begin{enumerate}
	\item Choose $\lambda_{01}(t)$, $\lambda_{12}(t \mid u)$ and some desired log rate ratio $\beta$.
	\item Choose some initial $\lambda_{02}^{(0)}(t)$.
	\item Calculate $r_{12}^{(0)}(t)$ from \eqref{eq:rate_calc} based on these.
\end{enumerate}
Then repeat until convergence:
\begin{enumerate}
	\setcounter{enumi}{3}
	\item Calculate $\lambda_{02}^{(j+1)}(t) = r_{12}^{(j)}(t) \mathrm{e}^{-\beta}$.
	\item Calculate $r_{12}^{(j+1)}(t)$ based on $\lambda_{01}(t), \lambda_{12}(t \mid u)$ and $\lambda_{02}^{(j)}(t)$.
\end{enumerate}
Step 4 ensures that $\lambda_{02}^{(j+1)}(t)$ is proportional to $r_{12}^{(j)}(t)$. However, changing $\lambda_{02}(t)$ also changes $r_{12}(t)$, so $r_{12}^{(j)}(t)$ is not the rate in the new model given by $\lambda_{01}(t), \lambda^{(j+1)}_{02}(t), \lambda_{12}(t \mid u)$. Therefore step 5 updates the rate to match again. This generally breaks the proportionality, but our hope is that the new pair $\lambda_{02}^{(j+1)}(t), r_{12}^{(j+1)}(t)$ is closer to proportionality than at step $j$.

\subsection{Extending to the frailty example}

When, as in the example in \Cref{sec:example}, the model is formulated conditional on an unmeasured frailty variable $Z$, the procedure in \Cref{sec:numerical_rates} needs to be slightly modified. The hazard conditional on frailty is
\begin{equation*}
    \begin{split}
        \lambda_{02}(t \mid Z) &= Z h_0(t) \\
        \lambda_{12}(t \mid U, Z) &= Z h_1(t),
    \end{split}
\end{equation*}
such that the model is now parametrized through the frailty distribution along with $h_0(t), h_1(t), \lambda_{01}(t)$ and $\beta$, with the marginal hazards $\lambda_{02}(t), \lambda_{12}(t \mid U)$ being given by \eqref{eq:marginal_hazard_Laplace}. With a Gamma frailty with mean 1 and variance $\theta$, the marginal hazards are
\begin{equation}
    \label{eq:marginal_hazard_Laplace_Gamma}
    \begin{split}
        \lambda_{02}(t) &= \frac{h_0(t)}{1 + \theta H_0(t)} \\
        \lambda_{12}(t \mid U) &= \frac{h_1(t)}{1 + \theta (H_0(U) + H_1(t \mid U))} ,
    \end{split}
\end{equation}
where $H_0(t) = \int_0^t h_0(s) \d s$ and $H_1(t \mid u) = \int_u^t h_1(s) \d s$. Note that the first equation implies that
\begin{equation}
    \label{eq:h_0}
    h_0(t) = \lambda_{02}(t) \e^{\theta \Lambda_{02}(t)}.
\end{equation}
With this, we extend the procedure in \Cref{sec:numerical_rates} as follows:
\begin{enumerate}
	\item Choose $\lambda_{01}(t)$, $h_1(t)$, Gamma frailty with variance $\theta$ and some desired log rate ratio $\beta$.
	\item Choose some initial $h_0^{(0)}(t)$.
	\item With $h_0^{(0)}(t)$ and $h_1(t)$, calculate $\lambda_{02}^{(0)}(t)$ and $\lambda_{12}^{(0)}(t \mid U)$ via \eqref{eq:marginal_hazard_Laplace_Gamma}.
    \item Calculate $r_{12}^{(0)}(t)$ from \eqref{eq:rate_calc} based on these.
\end{enumerate}
Then repeat until convergence:
\begin{enumerate}
	\setcounter{enumi}{4}
	\item Calculate $\lambda_{02}^{(j+1)}(t) = r_{12}^{(j)}(t) \mathrm{e}^{-\beta}$.
	\item Calculate $h_0^{(j+1)}(t)$ from $\lambda_{02}^{(j+1)}(t)$ via \eqref{eq:h_0}.
    \item With $h_0^{(j+1)}(t)$ and $h_1(t)$, calculate $\lambda_{12}^{(j+1)}(t \mid U)$ via \eqref{eq:marginal_hazard_Laplace_Gamma}.
    \item Calculate $r_{12}^{(j+1)}(t)$ from \eqref{eq:rate_calc} based on $\lambda_{02}^{(j+1)}(t)$ and $\lambda_{12}^{(j+1)}(t \mid U)$.
\end{enumerate}

\begin{figure}
	\centering
	\includegraphics[width=0.7\linewidth]{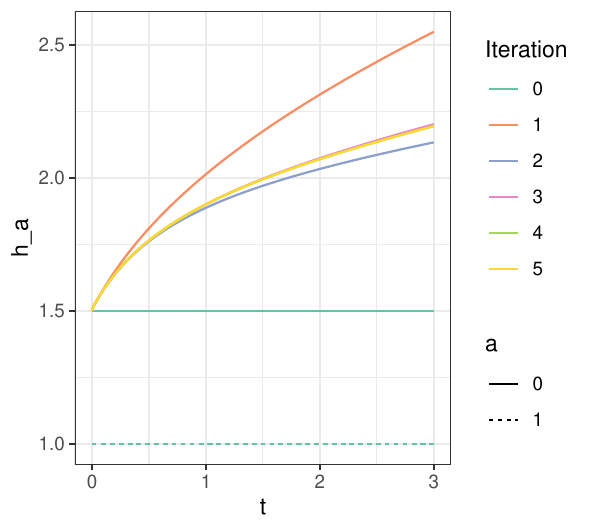}
	\caption{The functions $h_0^{(0)}(t)$ and $h_1(t)$, along with the obtained $h_0^{(j)}(t)$ for each iteration of the numeric scheme.}
	\label{fig:h_plot}
\end{figure}

\begin{figure}
	\centering
	\includegraphics[width=0.9\linewidth]{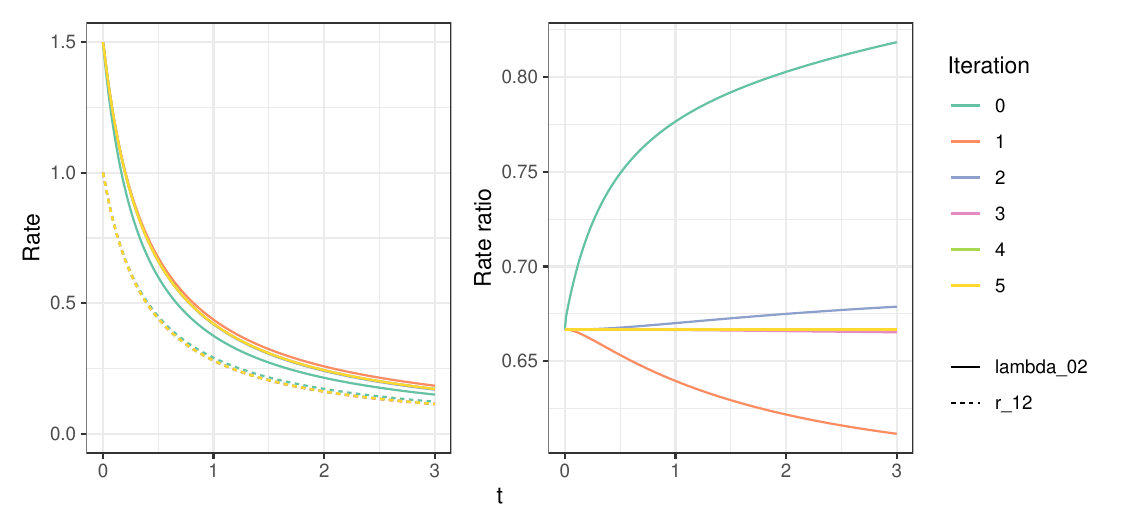}
	\caption{The obtained rates $r_{12}^{(j)}(t)$ and rate ratio $r_{12}^{[j)}(t) / \lambda_{02}^{(j)}(t)$ for each iteration of the numeric scheme.}
	\label{fig:prop_rate_iteration}
\end{figure}

We do not currently have a proof that the procedure converges. But numerically it seems to work. \Cref{fig:h_plot} shows the obtained $h_0(t)$ when applying the above procedure to the setup described in \Cref{sec:example} with the initial choice $h_0^{(0)}(t) = 3 / 2$. Similarly, \Cref{fig:prop_rate_iteration} shows obtained rates and rate ratio. After the third iteration, the rate ratio is essentially constant, and we have obtained a model which is Markov conditional on frailty, and where the observed rates are proportional, while the observed hazard is not Markov.

\section{Additive rates}
\label{sec:Additive rates}
\bigskip

\noindent
The Aalen additive rate model
$$
r(t \mid a(t-)) = \beta_0(t) + \beta_1(t) a(t-)
$$
is always correctly specified for binary $A(t)$, so we can estimate $B_j(t) = \int_0^t \beta_j(s) \d s$, $j = 0, 1$, consistently (and also the corresponding standard error using a sandwich estimator). The Aalen least squares estimator $\hat B(t)$ of $B(t)=\{B_0(t),B_1(t)\}^T$ is 
$$
\hat B(t)=\int_0^t\left \{Y^T(s)Y(s)\right\}^{-1}Y^T(s) \d N(s),
$$
where $N(t)=\{N_i(t)\}$ and the $i$'th row of the $n\times 2$ matrix $Y(t)$ is $Y_i(t)\{1,A_{i}(t)\}$. Doing the calculations gives 
\begin{align*}
	\d \hat B_0(t)&=\sum_iY_i(t)(1-A_{i}(t)) \d N_i(t)/Y_0(t),\\
	\d \hat B_1(t)&=\sum_iY_i(t)A_{i}(t) \d N_i(t)/Y_1(t)-\d \hat B_0(t),
\end{align*}
where $Y_0(t)=\sum_iY_i(t)(1-A_{i}(t))$ and $Y_1(t)=\sum_iY_i(t)A_{i}(t)$. If we let $\hat R_0(t)$ and $\hat R_0(t)$ denote the two Nelson-Aalen estimates corresponding to the two treatment groups, and using the technique from \citet{ snapinnIllustratingImpactTimeVarying2005} (ie. risk sets for the two treatment groups changes when subjects changes treatment) we then see that $ \d \hat B_0(t)=\d \hat R_0(t)$ and $ \d \hat B_1(t)=\d \hat R_1(t)-\d \hat R_0(t)$ so that the rate analysis corresponds to the extended Kaplan-Meier analysis of \citet{ snapinnIllustratingImpactTimeVarying2005}.

\end{document}